\documentclass[fleqn,usenatbib]{mnras}

\usepackage{newtxtext,newtxmath}
\usepackage[T1]{fontenc}
\usepackage{ae,aecompl}

\usepackage{graphicx}	% Including figure files
\usepackage{amsmath}	% Advanced maths commands
\usepackage{amssymb}	% Extra maths symbols

\usepackage{enumitem}
\usepackage{comment}
\usepackage{xcolor}

\hypersetup{draft}

\title[Cluster relaxation in DM simulations]{``Observing'' unrelaxed clusters in dark matter simulations}

\author[I. D. Roberts and L. C. Parker]{
Ian D. Roberts\thanks{E-mail: roberid@mcmaster.ca} and
Laura C. Parker
\\
Department of Physics and Astronomy, McMaster University, Hamilton ON L8S 4M1, Canada
}

\date{Accepted XXX. Received YYY; in original form ZZZ}

\pubyear{2018}

\begin{document}
\label{firstpage}
\pagerange{\pageref{firstpage}--\pageref{lastpage}}
\maketitle

\begin{abstract}
We present a detailed study of relaxed and unrelaxed galaxy clusters in a large dark-matter only simulation.  Recent work has demonstrated clear differences between the galaxy populations in clusters which have Gaussian velocity distributions (relaxed) compared to those that do not (unrelaxed).  To directly compare with observations, we identify unrelaxed clusters in the simulations using one-dimensional velocity distributions.  We show that non-Gaussian clusters have had recent major mergers and enhanced rates of galaxy infall relative to systems with Gaussian velocity profiles.  Furthermore, we find that the fraction of non-Gaussian clusters increases strongly with cluster mass and modestly with redshift.  For comparison, we also make use of three-dimensional information available in the simulations to explore the impact of projection on observational measurements.  Differences between Gaussian and non-Gaussian clusters are much stronger when three-dimensional information is considered, which demonstrates that the strength of observed trends with cluster dynamics are diluted by observed velocity information being limited to one line-of-sight.
\end{abstract}

\begin{keywords}
keyword1 -- keyword2 -- keyword3
\end{keywords}

%%%%%%%%%%%%%%%%%%%%%%%%%%%%%%%%%%%%%%%%%%%%%%%%%%

%%%%%%%%%%%%%%%%% BODY OF PAPER %%%%%%%%%%%%%%%%%%

\section{Introduction}

Galaxy clusters represent the largest virialized objects in the local Universe.  As such, galaxy clusters are important laboratories to understand the build up of massive galaxy environments as well as the influence of such extreme environments on satellite galaxy evolution.  However, observations of cluster substructures, both in the optical \citep[e.g.][]{dressler1988,girardi1997,flin2006,hou2012} and the X-ray \citep[e.g.][]{schuecker2001,jeltema2005,zhang2009} suggest that many clusters are not fully virialized.  These unrelaxed signatures are likely due to ongoing cluster formation, or a recent disruptive merger event.  The precise dynamical state of a given cluster can have significant impact on measured cluster properties as well as the evolution of galaxies members.  For instance, clusters which are not in dynamical equilibrium have measured velocity dispersions which are larger than the intrinsic cluster dispersion; this will lead to dynamical mass estimates which are biased high \citep[e.g.][]{old2018}.  Additionally, unrelaxed clusters may support an underdense intra-cluster medium (ICM) leading to low X-ray luminosities relative to relaxed systems \citep{popesso2007a, roberts2016, giles2017}.  This difference in ICM properties may have important implications for satellite quenching in clusters of different dynamical states.  Finally, if clusters appear unrelaxed due to ongoing formation and/or recent mergers, then the average time since infall for the satellite population should be short relative to relaxed clusters.  This will lead to satellite populations which have been exposed to the dense cluster environment for less time, and whose properties have therefore been comparatively less influenced by environment.
\par
Reliably identifying relaxed and unrelaxed clusters observationally is an active research topic, with two main approaches being commonly employed: 1. The use of X-ray observations; either photometrically by identifying unrelaxed clusters by the presence of a disturbed X-ray morphology, or spectroscopically by identifying relaxed clusters based on the presence of an X-ray cool-core \citep[e.g.][]{weissmann2013, nurgaliev2013}.  2. A dynamical analysis of cluster galaxies; for example through phase-space analyses \citep[e.g.][]{wojtak2013}, identifying galaxy substructures \citep[e.g.][]{hou2012}, or by classifying the shape of the member-galaxy velocity distribution \citep[e.g.][]{hou2009,ribeiro2013a}.  X-ray techniques are reliable, and relatively straight forward to apply, but require deep, high-resolution X-ray observations which are not available for most systems.  Dynamical approaches can be easily applied to large samples of groups and clusters from redshift surveys, but rely on high completeness and accurate determination of cluster membership.  Furthermore, both X-ray and dynamical approaches are complicated by the unavoidable fact that we lose information by observing galaxy clusters in projection.
\par
One of the simplest methods to classify cluster dynamical state is to examine the shape of the member-galaxy velocity distribution.  In \citet{roberts2018} we demonstrated that clusters with velocity distributions well fit by a Gaussian (G) have X-ray morphologies which are symmetric on average, whereas clusters with non-Gaussian (NG) velocity profiles show X-ray morphologies with significant asymmetries.  This suggests that the use of velocity distributions is a reliable way to determine cluster dynamical state.  Previous studies have found that NG clusters host an excess of blue, star-forming galaxies relative to G systems \citep{ribeiro2010,roberts2017}.  Velocity dispersion profiles (VDPs) also systematically differ between G and NG clusters, with relaxed clusters showing VDPs which decline with radius compared to rising or flat VDPs in unrelaxed clusters \citep{hou2009, costa2018, bilton2018}.
\par
A key missing ingredient in understanding G versus NG clusters is a detailed analysis of such systems in simulations.  Observations have established dependencies between galaxy properties and host-cluster dynamical state, and simulations give us access to key ``unobservables" such as cluster merger and infall history as well as 3-dimensional position and velocity information.  Given that cluster dynamics are dominated by the dark matter component we use a dark matter only simulation large enough to contain many galaxy cluster sized halos.  In this study we use the MultiDark Planck 2 simulation to study G and NG clusters.  We identify G and NG clusters from the simulation box using the same technique applied to observed clusters, which allows us to estimate unobservable properties such as time since last major merger and time since infall for satellites in clusters.  Furthermore, given detailed merger trees we can trace cluster halos back through time and constrain the timescales over which clusters appear NG.  Finally, we can gauge the effect of observational projection and determine whether observed trends are being diluted by misidentifying NG clusters in projection.   Again, given that we identify NG clusters using observational techniques, these properties are directly comparable to observed systems and can aid in interpreting observed differences between G and NG systems.
\par
The structure of this paper is as follows.  In Section~\ref{sec:methods} we introduce the simulation, our method for identifying galaxy-mass subhalos, and our method for identifying NG clusters.  In Section~\ref{sec:projection} we investigate the influence of projection on identifying G and NG clusters.  We explore the connection between NG clusters and recent major mergers as well as satellite time since infall in Sections~\ref{sec:MMtime} and \ref{sec:timeSinceInfall}, respectively.  In Section~\ref{sec:redshift} we investigate the evolution of NG clusters with redshift.  Finally, in Section~\ref{sec:discussion} we present and discuss the conclusions from this work. 

\section{Methods}
\label{sec:methods}

\subsection{MultiDark  Planck 2 simulation}

This paper uses data from the MultiDark Planck 2 (MDPL2, \citealt{prada2012, klypin2016}) simulation, a dark matter only simulation with a box-size of $(1000\,h^{-1}\,\mathrm{Mpc})^3$, assuming a flat $\Lambda$CDM cosmology with $h = 0.6777$, $\Omega_\Lambda = 0.692885$, $\Omega_m = 0.307115$, $\Omega_b = 0.048206$, $n_s = 0.96$, and $\sigma_8 = 0.8228$.  The simulation contains $3840^3$ particles with a mass resolution of $1.51 \times 10^9\,h^{-1}\,\mathrm{M_\odot}$, therefore resolving halos $>10^{11}\,h^{-1}\,\mathrm{M_\odot}$ with $\gtrsim 100$ particles.
\par
In each snapshot bound halos are identified with the phase-space friends-of-friends (FoF) algorithm \textsc{rockstar} \citep{behroozi2013a} and merger trees are generated with \textsc{consistent trees} \citep{behroozi2013b}.  Halo catalogues are output for 126 snapshots between $z = 15$ and $z = 0$.  Halo properties are calculated according to the virial overdensity, $\Delta_\mathrm{vir}(z)$, from \citet{bryan1998},
\begin{equation}
    \Delta_\mathrm{vir}(z) = 18\pi^2 + 82[\Omega(z)-1] - 39[\Omega(z)-1]^2
\end{equation}
\noindent
where, for a flat cosmology,
\begin{equation}
    \Omega(z) = \frac{\Omega_{m,0}(1+z)^3}{\Omega_{m,0}(1+z)^3 + \Omega_\Lambda}.
\end{equation}

\subsection{Identifying galaxy analogues}

In the dark-matter only simulation we identify subhalos that are of groups and clusters at $z=0$, and keep those with peak masses consistent with galaxies, following the procedure of \citet{joshi2016,joshi2017}.  In brief, starting with distinct halos at the top of the subhalo hierarchy, we select ``galaxy analogues" using the following criteria.

\begin{enumerate}[label={\arabic*.}, itemsep=0.5em, leftmargin=1em]
	\item %Ensure that galaxy analogues are well resolved and massive enough to be observed by large redshift surveys:\\[0.25em]
	If the peak halo mass, $M_\mathrm{peak}$, is $< 10^{11}\,h^{-1}\,\mathrm{M_\odot}$, the halo and its subsequent branches are not considered.
    
    \item %Ensure that galaxy analogues do not exceed the typical mass of high-mass galaxies:\\[0.25em]
    If $M_\mathrm{peak} > 10^{12.5}\,h^{-1}\,\mathrm{M_\odot}$, the halo is eliminated but each of its subhalos are put through criteria 1-4.
    
    \item %If the halo has no galaxy-mass subhalos, then accept it as a galaxy analogue:\\[0.25em]
    If $10^{11} < M_\mathrm{peak} < 10^{12.5}\,h^{-1}\,\mathrm{M_\odot}$ and the halo has no subhalos with $M_\mathrm{peak} > 10^{11}\,h^{-1}\,\mathrm{M_\odot}$, the halo is considered a galaxy analogue and its subsequent branches are eliminated.
    
    \item If $10^{11} < M_\mathrm{peak} < 10^{12.5}\,h^{-1}\,\mathrm{M_\odot}$ and the halo has at least one subhalo with $M_\mathrm{peak} > 10^{11}\,h^{-1}\,\mathrm{M_\odot}$, then the quantity $M_\mathrm{rem} = M_\mathrm{peak} - \sum M_{\mathrm{subhalo,}\,\mathrm{peak}}$ is considered.  If $10^{11} < M_\mathrm{rem} < 10^{12.5}\,h^{-1}\,\mathrm{M_\odot}$ then the halo is accepted as a galaxy analogue and each of its subhalos are put through criteria 1-4.
\end{enumerate}

\noindent
The above mass limits are chosen to correspond to the stellar mass range of galaxies in observational surveys ($M_\star \sim 10^9 - 10^{11}\,\mathrm{M_\odot}$, assuming a \citet{hudson2015} stellar-to-halo mass relationship.  The upper mass limit is chosen to avoid including group halos as part of our galaxy sample, however this cut may miss some massive central galaxies.  This is not a problem in this analysis because satellite galaxies are the primary tracers of the host cluster dynamical state.  It is also worth noting that halo finders struggle to accurately identify central substructure \citep[e.g.][]{knebe2011, joshi2016}, meaning that such massive subhalos sitting at the centre of the potential may be poorly identified.  In the $z = 0$ snapshot we identify $7\,308\,248$ galaxy analogues, and for brevity we will refer to galaxy analogues as ``galaxies" for the remainder of the paper\footnote{We emphasize that these ``galaxies'' are identified purely on dark matter content, with no consideration of stellar or gaseous components.}.

\subsection{Cluster dynamical states}
\label{sec:dynamical_state}

\begin{figure}
	\centering
    \includegraphics[width=\columnwidth]{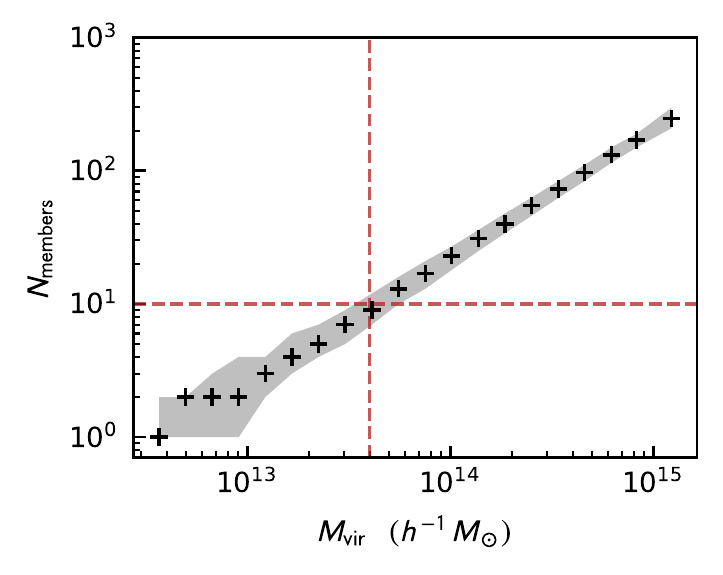}
    \caption{Median cluster membership (number of galaxies) as a function of cluster mass.  Shaded region corresponds to the 50 per cent (25th to 75th percentile) scatter.  Dashed lines mark $M_\mathrm{vir} = 4\times 10^{13}\,\mathrm{M_\odot}$ which is the cluster mass that corresponds to a median membership of $N_\mathrm{members} = 10$.  In our final sample we only include clusters with $M_\mathrm{vir} > 4\times 10^{13}\,\mathrm{M_\odot}$.}
    \label{fig:N_Mvir}
\end{figure}

\begin{figure*}
	\centering
    \includegraphics[width=0.9\textwidth]{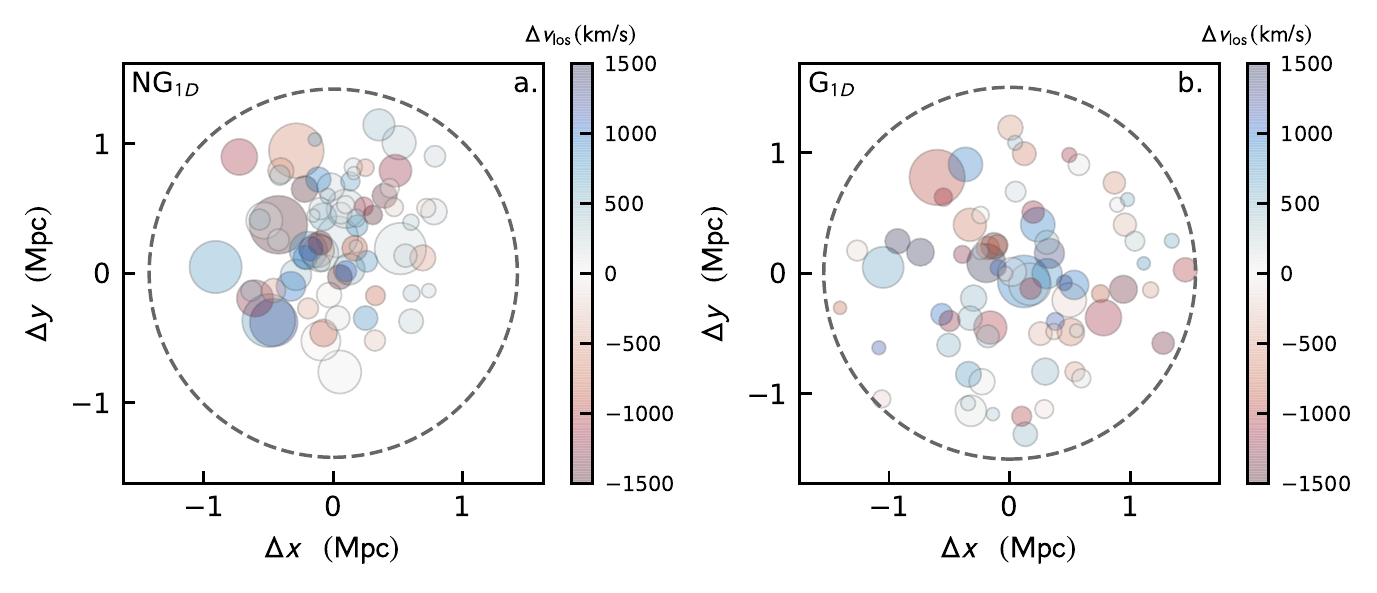}
    \caption{An example of a G (right) and NG (left) cluster identified by applying the Anderson-Darling test to cluster velocity distributions.  Circles correspond to the projected positions of cluster galaxy analogues and the dashed line marks the cluster virial radius.  The size of the circles are scaled according to the subhalo virial radii and they are coloured according to their $v_\mathrm{los}$ offset from the cluster centroid.}
    \label{fig:halo_image}
\end{figure*}

In order to characterize the velocity distribution of member galaxies with relative accuracy (while still maintaining a large sample of clusters) we only consider clusters with at least ten members \citep{hou2009,roberts2018}.  In this analysis we use galaxy analogues identified as cluster members by the \textsc{rockstar} halo finder.  However, our key findings remain the same if instead we assign cluster memberships using cuts in projected radius and 1D velocity dispersion (ie. more similar to observational memberships).  The membership cut restricts the sample size to $2\,300\,045$ galaxies in $101\,868$ clusters.  The host clusters (with 10+ members) range between $4.8 \times 10^{12} < M_\mathrm{vir} < 3.5 \times 10^{15}\,\mathrm{M_\odot}$.  Despite the large range in mass we refer to all host halos as clusters for simplicity.  When we explicitly consider dependencies on halo mass in the subsequent sections, we will refer to systems with $M_\mathrm{vir} < 10^{14}\,h^{-1}\,\mathrm{M_\odot}$ as \textit{low-mass clusters} and systems with $M_\mathrm{vir} \ge 10^{14}\,h^{-1}\,\mathrm{M_\odot}$ as \textit{high-mass clusters}.  The halo masses that we quote throughout the paper are the simulation halo masses from the \textsc{rockstar} catalogues.  We note that we have also performed our analysis using dynamical masses estimated from one-dimensional velocity dispersions (a common observational halo mass estimator) and find that our results are unchanged.  When estimating dynamical masses we find that measured velocity dispersions for high-mass NG clusters are enhanced by roughly 10 per cent relative to similar G clusters.  This is due to the fact that NG clusters are more dynamically disturbed, however we find that this small difference does not impact our results.
\par
In Fig.~\ref{fig:N_Mvir} we show median cluster membership (ie. the number of galaxies identified in each parent halo) as a function of parent halo mass.  For clusters with $M_\mathrm{vir} < 4 \times 10^{13}\,h^{-1}\,\mathrm{M_\odot}$ we note that the median cluster membership is less than 10.  This means that by selecting only clusters with 10+ members, we are biasing our sample at halo masses less than $4 \times 10^{13}\,h^{-1}\,\mathrm{M_\odot}$.  In our final sample we only include clusters with $M_\mathrm{vir} > 4 \times 10^{13}\,h^{-1}\,\mathrm{M_\odot}$ in order to avoid these potential biases.  This leaves a final sample consisting of $2\,000\,328$ galaxies in $77\,533$ clusters.
\par
To make direct comparisons to observations we consider galaxy positions and velocities in projection.  We project each cluster along a random axis mimicking the fact that real clusters are observed along a random line-of-sight (LOS).  We will refer to the two projected position axes as $\tilde{x}$ and $\tilde{y}$ and the projected velocity direction as $\tilde{z}$ for each cluster.  We stress that these are randomly projected axes for each cluster and in general do not correspond to the $x$, $y$, and $z$ coordinate axes of the simulation box.  We classify relaxed and unrelaxed clusters by considering the shape of the projected velocity distribution for member satellite galaxies \citep[e.g.][]{hou2009}.  This method is predicated on the notion that relaxed/dynamically old clusters will be characterized by velocity distributions which are close to Gaussian \citep{yahil1977,bird1993}.  To quantify the degree to which projected velocities are consistent with a Gaussian, we apply the Anderson-Darling (AD, \citealt{anderson1952}) normality test to the distribution of $v_{\tilde{z}}$ for each cluster in the sample.  Specifically, we consider

\begin{equation}
	v_\mathrm{los} = v_{\tilde{z}} - \bar{v}_{\tilde{z}}
\end{equation}
\noindent
where $v_{\tilde{z}}$ is the velocity in the random $\tilde{z}$-direction for each galaxy and $\bar{v}_{\tilde{z}}$ is the mean $v_{\tilde{z}}$ for galaxies in the cluster.
\par
The AD normality test is a non-parametric normality test which quantifies the distance between the cumulative distribution function (CDF) of the data and the CDF of a normal distribution.  This distance is parameterized by the AD statistic \citep{anderson1952,dagostino1986} given by:

\begin{equation}
	A^{*2} = A^2 \times (1.0 + 0.75/n + 2.25 / n^2)
\end{equation}
\noindent
where,
\begin{equation}
	A^{2} = -n - \frac{1}{n} \sum_{i=1}^n [2i - 1][\ln \Phi(x_i) + \ln (1 - \Phi(x_{n + 1 - i}))]
\end{equation}
\noindent
where $x_i$ are the length-$n$ ordered data and $\Phi(x_i)$ is the CDF of the Gaussian distribution.  A $p\text{-}value$ is then computed from the value of the AD statistic, $A^{*2}$, and following previous work \citep[e.g.][]{hou2009, roberts2017} we consider clusters with $p_{AD} < 0.05$ to be non-Gaussian in one-dimension ($\mathrm{NG_{1D}}$) and clusters with $p_{AD} \ge 0.05$ to be Gaussian in one dimension ($\mathrm{G_{1D}}$).  One important consideration is the fact that statistical normality tests such as the AD test will more readily detect subtle departures from normality when the sample size is large, due to the increasing statistical power of the test \citep[e.g.][]{razali2011}.  In our sample, for a given halo mass, the median cluster membership is the same for $\mathrm{G_{1D}}$ and $\mathrm{NG_{1D}}$ clusters (this is true at all halos masses we consider), therefore the fact that the statistical power of the AD test increases with sample size should not introduce any bias between our $\mathrm{G_{1D}}$ and $\mathrm{NG_{1D}}$ samples.  At $z = 0$, we find $72\,178$ $\mathrm{G_{1D}}$ clusters and $5\,355$ $\mathrm{NG_{1D}}$ clusters.  In Fig.~\ref{fig:halo_image} we show example clusters which we identify as $\mathrm{NG_{1D}}$ (left) and $\mathrm{G_{1D}}$ (right), each with masses of $M_\mathrm{vir} \simeq 4 \times 10^{14}\,h^{-1}\,\mathrm{M_\odot}$.  The dashed line corresponds to the virial radius of the cluster halo and the circles correspond to the projected ($\tilde{x}$, $\tilde{y}$) positions of member galaxies (sized according to the virial radius of the subhalo and coloured according to their velocity in the $\tilde{z}$-direction).

\section{Effects of line-of-sight projection}
\label{sec:projection}

\begin{figure}
	\centering
    \includegraphics[width=\columnwidth]{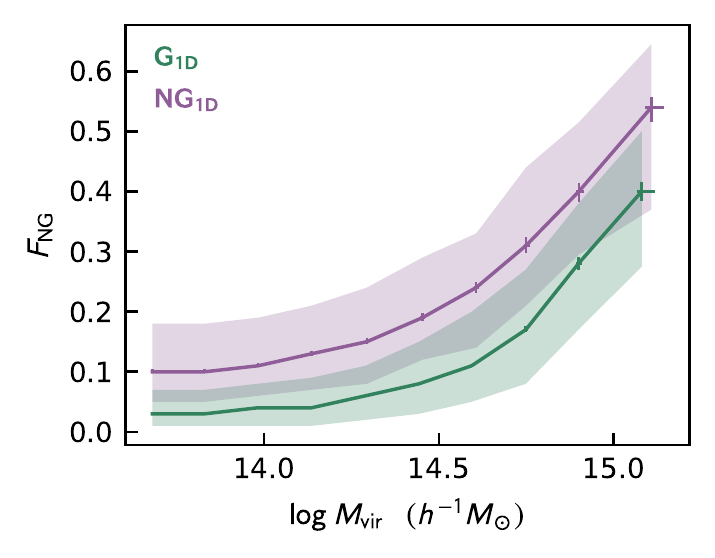}
    \caption{The fraction of 100 random projections along which a given cluster appears NG ($p_\mathrm{AD} < 0.05$) as a function of parent virial mass.  The purple line corresponds to $\mathrm{NG_{1D}}$ clusters and the green line corresponds to the $\mathrm{G_{1D}}$.  Error bars correspond to 68 per cent statistical errors estimated from the beta distribution following \citet{cameron2011}, shaded region corresponds to the 50 per cent (25th to 75th percentile) scatter.}
    \label{fig:Mvir_Funrel}
\end{figure}

Throughout this paper we will be analysing clusters which are classified as G or NG according to velocity distributions along a single LOS ($\mathrm{G_{1D}}$ and $\mathrm{NG_{1D}}$).  The advantage of using simulations is that we can also gauge the effects of misclassification due to projection and quantify the effect of this on the trends that we observe.  To do this, we develop an estimate of cluster dynamical state measured along many random lines-of-sight, as opposed to just one.  Using three-dimensional information allows a more robust understanding of the dynamical state of each cluster.  In order to quantify the effects of projection we reproject each cluster along a random LOS 100 times.  For the $i$th random projection we apply the AD test to the one-dimensional velocity distribution and classify the cluster as NG (along that specific LOS) if $p_{\mathrm{AD},i} < 0.05$.  For the 100 random projections we compute the fraction of realizations where the cluster is classified as NG, namely
\begin{equation}\label{eq:f_unrel}
    F_\mathrm{NG} = \frac{N(p_i < 0.05)}{N_\mathrm{tot}}
\end{equation}
\noindent
where $N(p_i < 0.05)$ is the number of random projections where the cluster is classified as NG and $N_\mathrm{tot}$ is the total number of random projections (100 in this case).  This fraction, $F_\mathrm{NG}$, is therefore a measure of how unrelaxed a given cluster appears along many lines-of-sight as opposed to just the one LOS we are limited to observationally.
\par
We can now compare $F_\mathrm{NG}$, which is measured for each cluster, for $\mathrm{G_{1D}}$ and $\mathrm{NG_{1D}}$ clusters.  In Fig.~\ref{fig:Mvir_Funrel} we plot median $F_\mathrm{NG}$ versus parent halo mass for clusters in the sample which are $\mathrm{G_{1D}}$ (green) and $\mathrm{NG_{1D}}$ (purple) at $z=0$.  The error bars correspond to 68 per cent statistical uncertainties and the shaded region shows the 50 per cent (25th to 75th percentiles) scatter.  Regardless of dynamical state, halo mass and $F_\mathrm{NG}$ are strongly correlated, with $F_\mathrm{NG}$ increasing towards high halo masses.  This reflects the fact that high-mass clusters are inherently less virialized than lower mass systems \citep[e.g.][]{press1974}.  The median $F_\mathrm{NG}$ is systematically larger for $\mathrm{NG_{1D}}$ clusters compared to $\mathrm{G_{1D}}$ clusters at all halo masses.  At low halo masses $F_\mathrm{NG}$ is small, $\sim$0 for $\mathrm{G_{1D}}$ clusters and $\sim$0.1 for $\mathrm{NG_{1D}}$ clusters.  At these masses clusters are classified as G along most lines-of-sight, even those clusters that were classified as NG along one random LOS (purple), which have $F_\mathrm{NG}\simeq0.1$. Fig.~\ref{fig:Mvir_Funrel} demonstrates the impurity that can be introduced when restricted to observing along a single LOS; there is always the chance that the observed LOS may not be reflective of the dynamics of the cluster as a whole.  On the high mass end $F_\mathrm{NG}$ is much larger, $\sim$0.3-0.4 for $\mathrm{G_{1D}}$ clusters and $\sim$0.5 for $\mathrm{NG_{1D}}$ clusters.  While on the low-mass end the sample of $\mathrm{NG_{1D}}$ clusters is likely contaminated by clusters with relatively relaxed dynamics, on the high-mass end the converse is true.  The high impurity for low-mass systems in Fig.~\ref{fig:Mvir_Funrel} suggests that the AD test in one-dimension struggles to identify truly unrelaxed systems at the low end of our mass range.  This could be due to the fact that these systems have fewer member galaxies and therefore with a small number of dynamical tracers we may be under-sampling the underlying halo velocity profile.  The fact that $F_\mathrm{NG} \sim 0.3{\text -}0.4$ for high-mass $\mathrm{G_{1D}}$ clusters suggests that the $\mathrm{G_{1D}}$ sample contains some clusters which show complex dynamical states along many lines-of-sight.  Therefore despite the fact that high-mass $\mathrm{G_{1D}}$ clusters appear relaxed along a single, random LOS many of these clusters may look much less relaxed with three-dimensional information.
\par
Regardless of the value of $F_\mathrm{NG}$, the fact that $F_\mathrm{NG}$ is systematically larger for $\mathrm{NG_{1D}}$ clusters compared to $\mathrm{G_{1D}}$ clusters demonstrates that the AD test is selecting $\mathrm{NG_{1D}}$ clusters which are inherently less relaxed than their $\mathrm{G_{1D}}$ counterparts, at all halo masses.  Therefore this method works on average when applied to a large sample of clusters, but not necessarily for an individual system.

\section{Mass-matched cluster sample}
\label{sec:mass-match}

\begin{figure*}
	\centering
    \includegraphics[width=0.9\textwidth]{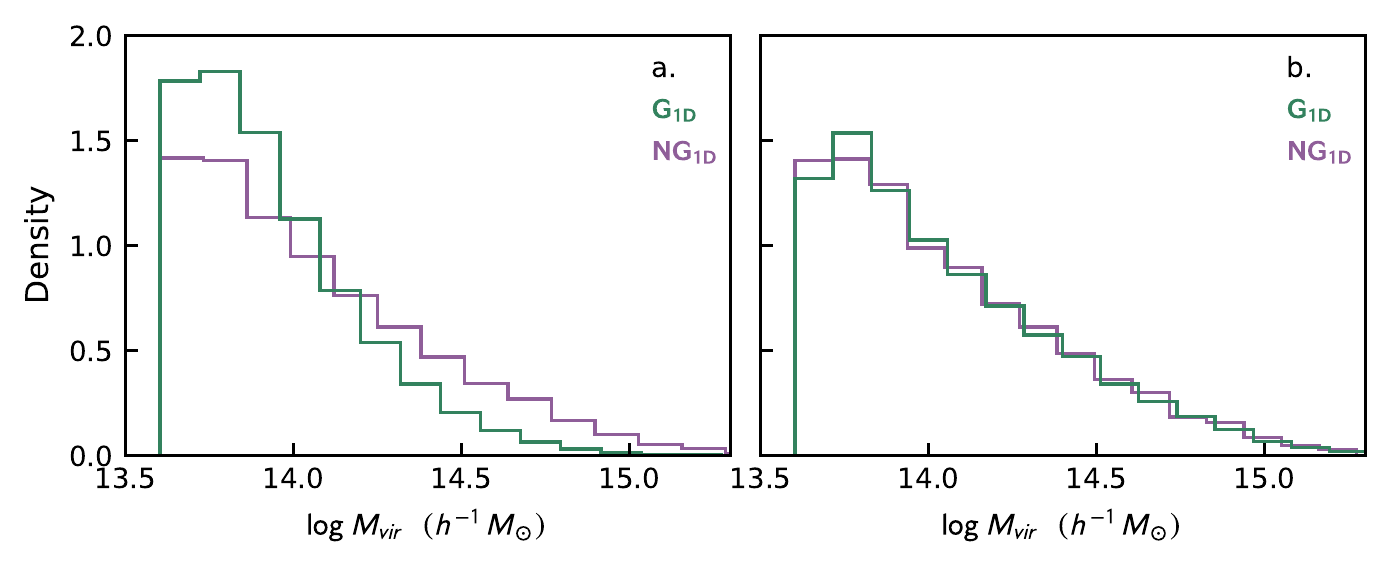}
    \caption{Virial mass distributions for $\mathrm{G_{1D}}$ (green) and $\mathrm{NG_{1D}}$ (purple) clusters in the sample.  \textit{Left:}  Distributions for the total sample.  \textit{Right:}  Distributions for the mass-matched sample.}
    \label{fig:mass_dist}
\end{figure*}

Fig.~\ref{fig:Mvir_Funrel} shows a clear dependence between cluster mass and the shape of the velocity profile, namely high mass clusters are far more likely to be classified as NG.  This is further illustrated in Fig.~\ref{fig:mass_dist}a where we show the distribution of virial mass for $\mathrm{G_{1D}}$ (green) and $\mathrm{NG_{1D}}$ (purple) clusters.  It is apparent that there is a small excess of $\mathrm{NG_{1D}}$ clusters at the highest cluster masses.  This excess has also been previously reported in samples of observed clusters \citep{roberts2017, decarvalho2017}.  Given this dependence of classified dynamical state on cluster mass, it is important to mass-match the $\mathrm{G_{1D}}$ and $\mathrm{NG_{1D}}$ samples to ensure that any differences between $\mathrm{G_{1D}}$ and $\mathrm{NG_{1D}}$ clusters are not resulting from different cluster mass distributions.
\par
In order to construct a mass-matched sample of $\mathrm{G_{1D}}$ and $\mathrm{NG_{1D}}$ clusters, for each $\mathrm{NG_{1D}}$ cluster we select 10 $\mathrm{G_{1D}}$ clusters with have virial masses within 0.1 dex of the $\mathrm{NG_{1D}}$ cluster.  The 10:1 is chosen to roughly match the ratio of $\mathrm{G_{1D}}$ to $\mathrm{NG_{1D}}$ clusters identified by the AD test (see section~\ref{sec:dynamical_state}).  In Fig.~\ref{fig:mass_dist}b we now plot the virial mass distributions for $\mathrm{G_{1D}}$ and $\mathrm{NG_{1D}}$ clusters in the mass-matched sample, clearly showing that the mass distributions of the two subsamples are now well matched.  For the remainder of the paper, any results comparing properties of $\mathrm{G_{1D}}$ and $\mathrm{NG_{1D}}$ clusters will show trends for both the original sample of $\mathrm{G_{1D}}$ and $\mathrm{NG_{1D}}$ clusters as well as the mass-matched sample.  The differences seen between $\mathrm{G_{1D}}$ $\mathrm{NG_{1D}}$ clusters cannot be explained by differences in the cluster mass distributions (see Figs~\ref{fig:majormerger} and \ref{fig:infallTime}).

\section{Time since last major merger}
\label{sec:MMtime}

\begin{figure*}
	\centering
    \includegraphics[width=0.9\textwidth]{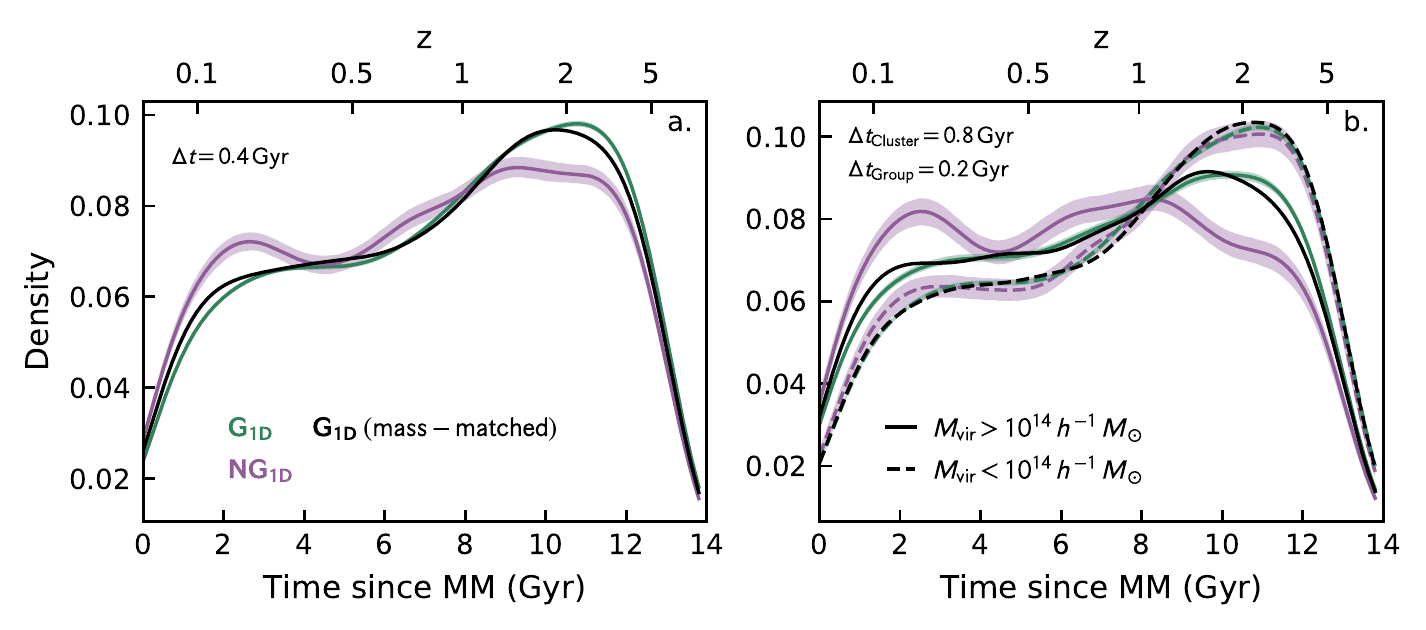}
    \caption{Distribution of time since last major merger for $\mathrm{G_{1D}}$ (green) and $\mathrm{NG_{1D}}$ (purple) clusters.  The solid black line corresponds to the $\mathrm{G_{1D}}$ sample which is mass-matched to the $\mathrm{NG_{1D}}$ sample (see section~\ref{sec:mass-match}).  Distributions are generated with a Gaussian kernel density estimation.  Shaded regions show the 68 per cent confidence region from 1000 random bootstrap re-samplings.  \textit{Left:} Distributions for the entire sample.  \textit{Right:}  Distributions split into low-mass ($M_\mathrm{vir} < 10^{14}\,h^{-1}\,\mathrm{M_\odot}$) and high-mass clusters ($M_\mathrm{vir} > 10^{14}\,h^{-1}\,\mathrm{M_\odot}$).  The median time difference between $\mathrm{G_{1D}}$ and $\mathrm{NG_{1D}}$, $\Delta t$, is shown for each sample.}
    \label{fig:majormerger}
\end{figure*}

A useful proxy for the dynamical age of a cluster halo is the time since last major merger (MM) \citep[e.g.][]{rowley2004}.  The scale-factor of the last MM for each halo is given in the \textsc{rockstar} halo catalogues (defining a major merger to have a mass ratio greater than 0.3), which is easily converted to a time since last major merger given our assumed cosmology.  In Fig.~\ref{fig:majormerger}a we show the kernel density distribution of time since last MM for $\mathrm{G_{1D}}$ (green) and $\mathrm{NG_{1D}}$ (purple) clusters.  The distributions for both $\mathrm{G_{1D}}$ and $\mathrm{NG_{1D}}$ clusters peak at early times ($t_\mathrm{lookback} \sim 10-12\,\mathrm{Gyr}$) corresponding to early cluster assembly, with the $\mathrm{G_{1D}}$ clusters showing an excess relative to $\mathrm{NG_{1D}}$ systems.  At late times ($t_\mathrm{lookback} \lesssim 4\,\mathrm{Gyr}$) the distribution of $\mathrm{NG_{1D}}$ systems shows a secondary peak corresponding to late-time MMs which is smaller for $\mathrm{G_{1D}}$ clusters.  As a whole, the median time since last MM is $0.4\,\mathrm{Gyr}$ shorter for $\mathrm{NG_{1D}}$ clusters compared to $\mathrm{G_{1D}}$ systems.  Therefore the AD test for cluster dynamics is senstive to physical differences in cluster merger history, namely $\mathrm{NG_{1D}}$ systems have preferentially short times since MM.  This difference is subtle but systematic, suggesting that the AD test applied to large samples of groups and clusters can identify statistical differences in merger history.
\par
In Fig.~\ref{fig:majormerger}b we show the same distributions, but now divided into low-mass (dashed, $M_\mathrm{vir} < 10^{14}\,h^{-1}\,\mathrm{M_\odot}$) and high-mass  (solid, $M_\mathrm{vir} \ge 10^{14}\,h^{-1}\,\mathrm{M_\odot}$) clusters.  Nearly all of the difference seen in Fig.~\ref{fig:majormerger}a is driven by the high-mass clusters, as low-mass $\mathrm{G_{1D}}$ and $\mathrm{NG_{1D}}$ clusters have virtually identical time since last MM distributions.  When considering only the high-mass clusters, the difference from Fig.~\ref{fig:majormerger}a becomes larger with a median difference between $\mathrm{G_{1D}}$ and $\mathrm{NG_{1D}}$ of $0.8\,\mathrm{Gyr}$.  The little difference in time since MM distributions for low-mass $\mathrm{G_{1D}}$ and $\mathrm{NG_{1D}}$ clusters suggests that the AD test is not identifying clear physical differences (at least in terms of MMs) for low-mass clusters like it is for high-mass clusters.  Indeed, the fraction of low-mass $\mathrm{NG_{1D}}$ clusters ($M_\mathrm{vir} < 10^{14}\,h^{-1}\,\mathrm{M_\odot}$) is only 5.5 per cent.  Given that the fraction of low-mass $\mathrm{NG_{1D}}$ clusters is close to the $p{\text -}value$ used to to identify $\mathrm{NG_{1D}}$ systems ($p_\mathrm{AD} = 0.05$), we can't rule out that many of the low-mass clusters that we identify as $\mathrm{NG_{1D}}$ are false-positives with intrinsic velocity distributions drawn from a Gaussian.  Indeed, in Fig.~\ref{fig:Mvir_Funrel} we show that low-mass clusters which are classified as $\mathrm{NG_{1D}}$ actually appear G along most lines-of-sight.

\subsection{Increasing the purity of the unrelaxed sample}

\begin{figure}
	\centering
    \includegraphics[width=\columnwidth]{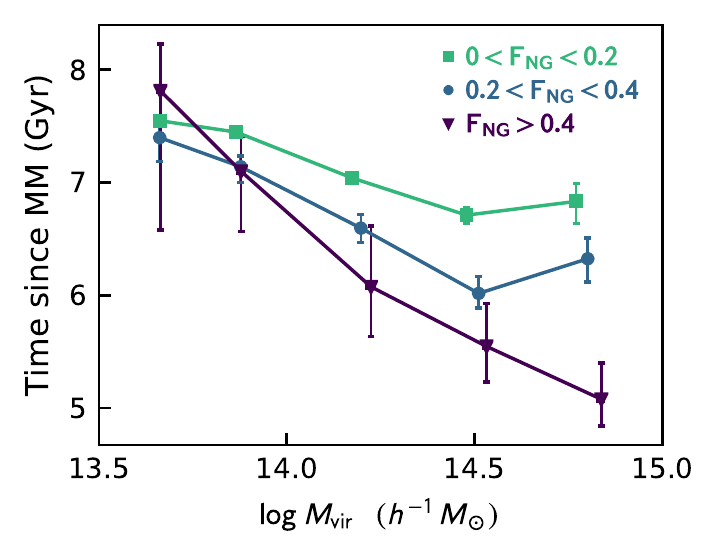}
    \caption{Time since major merger as a function of cluster mass, in bins of $F_\mathrm{NG}$ (see Equation~\ref{eq:f_unrel}).  Error bars are 68 per cent uncertainties estimated non-parametrically as: |16/84th percentile - median| / sqrt(N).}
    \label{fig:MMtime_pFrac}
\end{figure}

To construct a sample of NG clusters with higher purity we use $F_\mathrm{NG}$ (see Section~\ref{sec:projection}, Equation~\ref{eq:f_unrel}).  A reminder that $F_\mathrm{NG}$ corresponds to the fraction of random projections, for a given cluster, along which the cluster is classified as NG.  Therefore a sample of clusters with large values of $F_\mathrm{NG}$ will be a sample of unrelaxed clusters with relatively high purity.  If the AD test is identifying physical differences between clusters classified as G and NG then we expect the properties of relaxed and unrelaxed clusters to differ more strongly as $F_\mathrm{NG}$ increases.  In other words, the differences between relaxed and unrelaxed samples should increase as the purity of the unrelaxed sample increases.  In Fig.~\ref{fig:MMtime_pFrac} we plot the average time since MM as a function of cluster mass, for different bins of $F_\mathrm{NG}$.  Given the mass dependence of $F_\mathrm{NG}$ it is important to compare $F_\mathrm{NG}$ at fixed cluster mass.  The errorbars in Fig.~\ref{fig:MMtime_pFrac} are computed non-parametrically as: |16/84th percentile - median| / sqrt(N).  Fig.~\ref{fig:MMtime_pFrac} shows different trends for different cluster masses.  For low-mass clusters there is no trend between time since MM and $F_\mathrm{NG}$, which may be related to the fact that low-mass clusters have relatively few recent MMs (see Fig.~\ref{fig:majormerger}b and Fig.~\ref{fig:MMfrac}b).  Furthermore, the frequency of MMs over the entire cluster lifetime is lower for low-mass clusters compared to high-mass systems (plot not shown).  On the other hand, for high-mass clusters there is a clear anticorrelation between time since MM and $F_\mathrm{NG}$.  For high-mass clusters, a large fraction of projections that show NG dynamics corresponds to relatively short time since MM.  The difference in time since MM between the smallest and largest values of  $F_\mathrm{NG}$ ranges from  $\sim 1-2\,\mathrm{Gyr}$ for the higher mass clusters.  This difference highlights the inherent information lost when restricted to observing along a single line-of-sight.  While a systematic difference in time since MM between $\mathrm{G_{1D}}$ and $\mathrm{NG_{1D}}$ clusters is seen for the one-dimensional case (Fig.~\ref{fig:majormerger}), when considering a more pure sample of unrelaxed clusters (high values of $F_\mathrm{NG}$) the difference is strongly enhanced (Fig.~\ref{fig:MMtime_pFrac}).

\subsection{Recent merger fractions}

\begin{figure}
	\centering
    \includegraphics[width=\columnwidth]{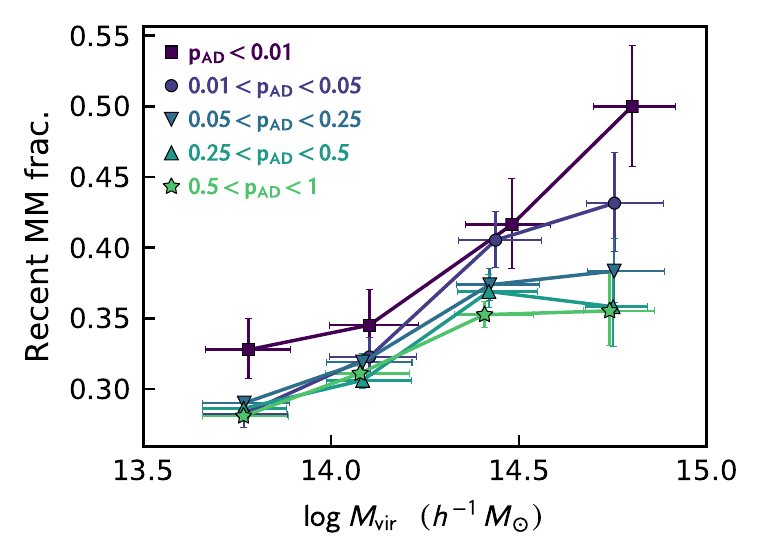}
    \caption{Fraction of $z=0$ clusters with a major merger within the past 5 Gyr as a function of cluster mass, for bins of Anderson-Darling p-value.  The errors correspond to 68 per cent statistical errors estimated from the beta distribution following \citet{cameron2011}.}
    \label{fig:MMfrac}
\end{figure}

Given that the distribution of time since MM for $\mathrm{NG_{1D}}$ clusters in this sample appears bimodal, it is natural to divide the population into two classes: clusters which have experienced a recent MM, and those which have not.  Based on the distributions in Fig.~\ref{fig:majormerger} we define a recent MM to be a MM within the last $5\,\mathrm{Gyr}$, though our results are not sensitive to the specific dividing line that we choose.  In Fig.~\ref{fig:MMfrac} we show the fraction of clusters that have experienced a recent MM (time since MM < 5 Gyr) as a function of cluster virial mass, for different bins of AD \textit{p-value}.  For all values of AD \textit{p-value} there is a correlation between recent MM fraction and cluster mass.  Recent MM fraction increases most strongly with cluster mass for the low values of AD \textit{p-value}, specifically for clusters which we classify as $\mathrm{NG_{1D}}$ ($p_{AD} < 0.05$).  Furthermore, at fixed cluster mass the recent MM fraction increases with decreasing AD \textit{p-value}.  This is most obvious at the high-mass end where the recent MM fractions are clearly highest for the smallest \textit{p-values}.  In other words, the fraction of clusters which have experienced a recent MM is highest for systems which appear very dynamically disturbed.

\section{Satellite time since infall}
\label{sec:timeSinceInfall}

\begin{figure}
	\centering
    \includegraphics[width=\columnwidth]{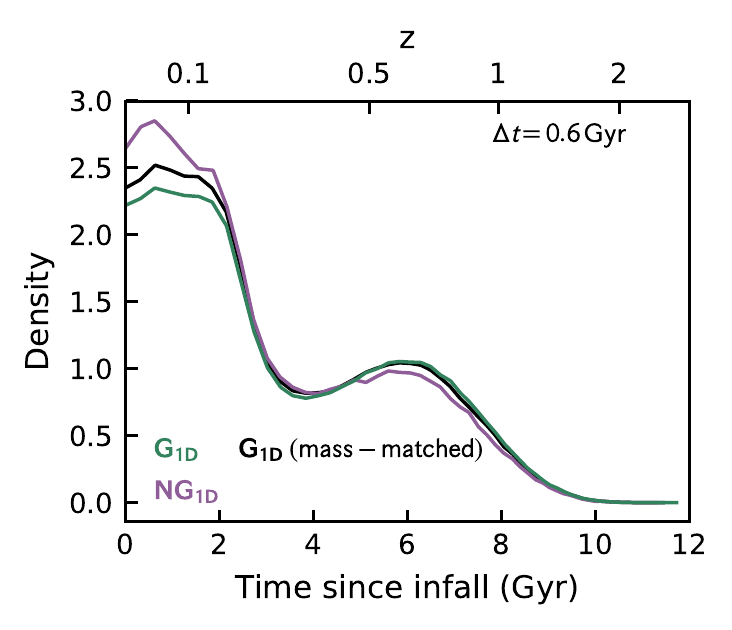}
    \caption{Distribution of time since infall for galaxies in $\mathrm{G_{1D}}$ (green) and $\mathrm{NG_{1D}}$ (purple) clusters.  The solid black line corresponds to the $\mathrm{G_{1D}}$ sample which is mass-matched to the $\mathrm{NG_{1D}}$ sample (see section~\ref{sec:mass-match}).  The median time difference between $\mathrm{G_{1D}}$ and $\mathrm{NG_{1D}}$ clusters, $\Delta t$, is shown.}
    \label{fig:infallTime}
\end{figure}

We now consider the infall history of galaxies onto their present-day parent halos.  Infall history is related to time since last MM, since mergers are a source of newly infalling satellites, however clusters are also continuously accreting new satellites which are not associated with rare MMs.  Time since infall for satellite galaxies is particularly interesting for exploring environmental quenching of star formation in galaxies, as observed quenched fractions are reproduced well by models which directly tie quenching to an infall time threshold \citep[e.g.][]{haines2015}.  It is possible that differences in observed quenched fractions between G and NG clusters \citep[e.g.][]{roberts2017} may be directly related to differences in time since infall.
\par
We derive time since infall onto the current parent halo for each galaxy by tracing the galaxy's most-massive progenitor (MMP) back through the merger trees\footnote{https://ytree.readthedocs.io/}.  We consider infall to be the first time that a MMP of a galaxy becomes a subhalo of a MMP of the galaxy's present day parent halo.  In Fig.~\ref{fig:infallTime} we plot the time since infall distributions for the $\mathrm{G_{1D}}$ and $\mathrm{NG_{1D}}$ samples.  Galaxies in $\mathrm{NG_{1D}}$ systems have systematically shorter times since infall, with a median difference of $0.6\,\mathrm{Gyr}$.  Again, the difference between $\mathrm{G_{1D}}$ and $\mathrm{NG_{1D}}$ clusters is subtle but systematic. Similarly, observational studies have reported enhanced accretion in NG clusters relative to G clusters \citep{decarvalho2019}.  In both $\mathrm{G_{1D}}$ and $\mathrm{NG_{1D}}$ clusters, recent accretion dominates and the peak in time since infall occurs within the last $2\,\mathrm{Gyr}$.
\par
The bimodal shape seen in Fig.~\ref{fig:infallTime} is likely driven by backsplashing galaxies \citep{yun2019}.  Membership is restricted to those galaxies which are within the virial radius of the parent halo at $z=0$, therefore any galaxies which have made a pericentric passage and then ``backsplashed" beyond the virial radius will not be included as members.  The characteristic timescale required for a galaxy to infall, make a pericentric passage, and then backsplash beyond the virial radius is of order $\sim$few Gyr \citep[e.g.][]{oman2013}.  Therefore the deficit of satellites which have time since infall of 3-4 Gyr is likely related to those satellites backsplashing at $z=0$ and not being identified as members.  The distributions in Fig.~\ref{fig:infallTime} also do not account for satellites which were once members but have since been destroyed by tidal interactions or have merged with another galaxy.

\subsection{Increasing the purity of the unrelaxed sample}

\begin{figure*}
	\centering
    \includegraphics[width=0.9\textwidth]{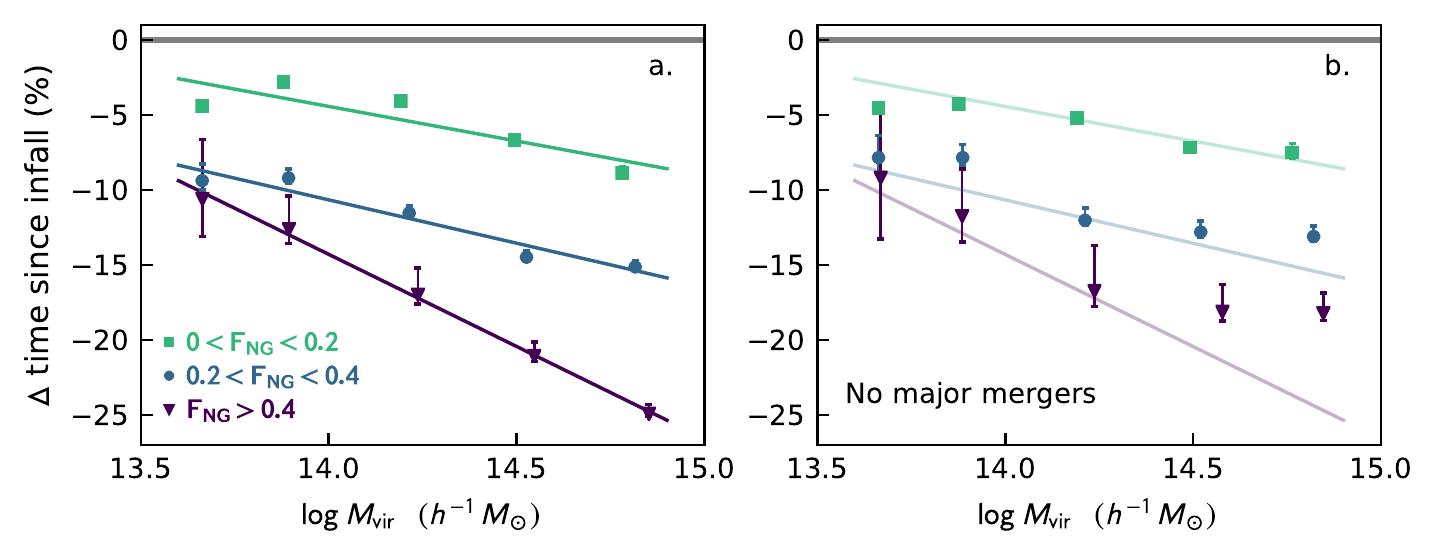}
    \caption{Percent change in time since infall (relative to ``purely" G systems, $F_\mathrm{NG}=0$) as a function of cluster mass, for bins of $F_\mathrm{NG}$ (see Equation~\ref{eq:f_unrel}).  Solid lines correspond to linear fits to the data in panel a.  Error bars are 68 per cent uncertainties estimated non-parametrically as: |16/84th percentile - median| / sqrt(N).  \textit{Left:} Time since infall for all clusters in the sample.  \textit{Right:}  Time since infall since the last major merger for galaxies in clusters that have not had a major merger in the past 8 Gyr, separating the contribution of infall from major mergers from panel a.}
    \label{fig:infallTime_pFrac}
\end{figure*}

Analogously to Fig.~\ref{fig:MMtime_pFrac}, we now investigate average time since infall as a function of cluster mass, for bins of the fraction of random projections along which a cluster is classified as NG, $F_\mathrm{NG}$, in Fig.~\ref{fig:infallTime_pFrac}a.  We normalize the y-axis such that we are plotting the percentage change in time since infall relative to clusters with $F_\mathrm{NG}=0$ (relaxed in all random projections).  Namely,
\begin{equation}
    \Delta \mathrm{time\,since\,infall} = 100 \times \frac{t_\mathrm{since\,infall}(F_\mathrm{NG}) - t_\mathrm{since\,infall}(F_\mathrm{NG}=0)}{t_\mathrm{since\,infall}(F_\mathrm{NG}=0)}
\end{equation}
\noindent
where $t_\mathrm{since\,infall}(F_\mathrm{NG})$ is the average time since infall for galaxies as a function of $F_\mathrm{NG}$.
The shaded horizontal line in Fig.~\ref{fig:infallTime_pFrac} corresponds to the average time since infall for satellites of halos with $F_\mathrm{NG} = 0$, which in this case is $\Delta \mathrm{time\,since\,infall} = 0$ by construction.  For high-mass clusters we see a qualitatively similar trend to Fig.~\ref{fig:MMtime_pFrac} (time since MM), where time since infall decreases with increasing $F_\mathrm{NG}$.  High-mass clusters which appear dynamically unrelaxed along many lines-of-sight host satellites which have recently become members.  The trend for low-mass systems is clearly different when comparing satellite time since infall to time since MM.  Whereas no strong trend is seen between time since MM and $F_\mathrm{NG}$ (see Fig.~\ref{fig:MMtime_pFrac}), a clear trend is apparent between satellite time since infall and $F_\mathrm{NG}$.  Specifically, time since infall decreases with increasing $F_\mathrm{NG}$.  While the trend is weaker in low-mass clusters compared to high-mass clusters, a clear anti-correlation is present.
\par
The combination of Fig.~\ref{fig:MMtime_pFrac} and Fig.~\ref{fig:infallTime_pFrac} suggest that for low-mass systems the AD test is likely tracing satellite time since infall more than time since MM.  Infalling satellites onto clusters can be sourced through major mergers or more continuous accretion of small groups and individual galaxies.  Given the lack of dependence between $F_\mathrm{NG}$ and time since MM for low-mass clusters (see Fig.~\ref{fig:MMtime_pFrac}), it appears that the AD test is tracing continuous accretion as opposed to MMs for these lower mass systems.  Since low-mass clusters also have low galaxy memberships, it may be that this continuous accretion can have a larger impact on the dynamics of the host system.
\par
Major mergers will always facilitate the infall of new satellite galaxies onto a cluster, therefore the trends that we see in Fig.~\ref{fig:infallTime_pFrac}a are a superposition of infall associated with MMs as well as continuous accretion.  We separate the effect due to infall from continuous accretion versus MMs by selecting a subset of clusters which have not experienced any recent MMs, and therefore any recent accretion of satellites onto these systems will be driven by minor mergers and isolated accretion.  Specifically, we select all clusters which have not experienced a MM in the last 8 Gyr and consider only satellite infall occurring after the last MM.  This cut completely excludes the MM peak at late times (see Fig.~\ref{fig:majormerger}a), and we note that these results are not particularly sensitive to the specific dividing line chosen.
\par
In Fig.~\ref{fig:infallTime_pFrac}b we show $\Delta \mathrm{time\,since\,infall}$ versus $F_\mathrm{NG}$ for galaxies which have infallen since the last MM onto clusters that have not had a MM for at least 8 Gyr.  Therefore we have effectively removed the contribution from accretion via MMs from Fig.~\ref{fig:infallTime_pFrac}a.  To guide the eye we also show solid lines corresponding to weighted least-squares linear fits to the data in panel a.  For both low-mass and high-mass clusters there is still a residual trend between time since infall and $F_\mathrm{NG}$ suggesting the the AD test is sensitive to physical differences in infall history, even in the absence of MMs.  For the lowest-mass clusters there is little difference between panels a and b, which is consistent with the lack of clear correlation between $F_\mathrm{NG}$ and time since MM which was previously shown (Fig.~\ref{fig:MMtime_pFrac}).  As expected, removing the contribution from MM accretion has only a small effect for low-mass systems.  Conversely, the trend for high-mass clusters differs between Fig.~\ref{fig:infallTime_pFrac}a and Fig.~\ref{fig:infallTime_pFrac}b (especially for large values of $F_\mathrm{NG}$), showing that the both MMs and accretion contribute to the trend for high-mass systems.  Upon removing the MM contribution the trend between time since infall and $F_\mathrm{NG}$ becomes flatter.  This is especially clear when comparing the $F_\mathrm{NG} > 0.4$ best-fit line from panel a (solid purple line in panel b) to the $F_\mathrm{NG} > 0.4$ data points in panel b.  These results suggest that for massive clusters the AD test traces dynamical disruptions from major mergers and continuous accretion, whereas for lower-mass clusters the AD test seems to be primarily sensitive to continuous accretion and not MMs.

\section{Redshift evolution}
\label{sec:redshift}

\begin{figure}
	\centering
    \includegraphics[width=\columnwidth]{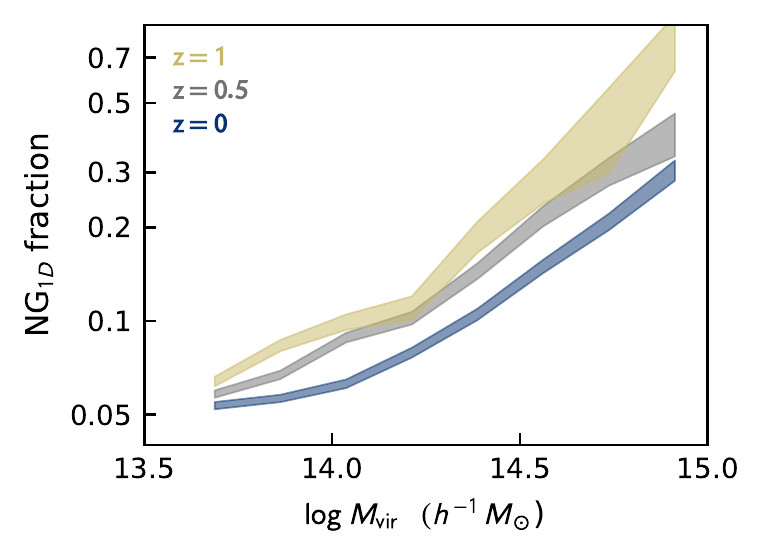}
    \caption{The fraction of $\mathrm{NG_{1D}}$ clusters versus halo mass for three different redshift snapshots.  The shaded regions correspond to 68 per cent statistical errors estimated from the beta distribution following \citet{cameron2011}.}
    \label{fig:NGfrac_z}
\end{figure}

\subsection{Fraction of unrelaxed clusters}

In Fig.~\ref{fig:NGfrac_z} we show the evolution of the fraction of clusters classified as $\mathrm{NG_{1D}}$ ($p_{AD} < 0.05$ along one random line-of-sight) as a function of halo mass and redshift, for three redshift snapshots ($z = 0, 0.5, 1$).  Fig.~\ref{fig:NGfrac_z} reveals two clear trends.  First, at all redshifts the fraction of $\mathrm{NG_{1D}}$ clusters increases with halo mass, and second, at fixed halo mass the fraction of $\mathrm{NG_{1D}}$ clusters increases modestly with redshift.  Both the trend with redshift and the trend with halo mass can be explained through simple virialization.  At all redshifts high-mass clusters are, on average, less virialized than lower mass halos, leading to more $\mathrm{NG_{1D}}$ clusters at high halo mass; and at all masses halos are, on average, less virialized at earlier epochs compared to the present day.
\par
These trends with halo mass and redshift are in qualitative agreement with observations of G and NG clusters.  Observations have shown that the proportion of NG to G systems increases at high halo mass \citep{ribeiro2013b, roberts2017, decarvalho2017}.  As well, observations of G and NG systems at different redshifts have demonstrated that the fraction of NG clusters tends to increase with redshift \citep{hou2013}.

\subsection{How long have $z = 0$ clusters appeared unrelaxed?}

\begin{figure*}
	\centering
    \includegraphics[width=0.9\textwidth]{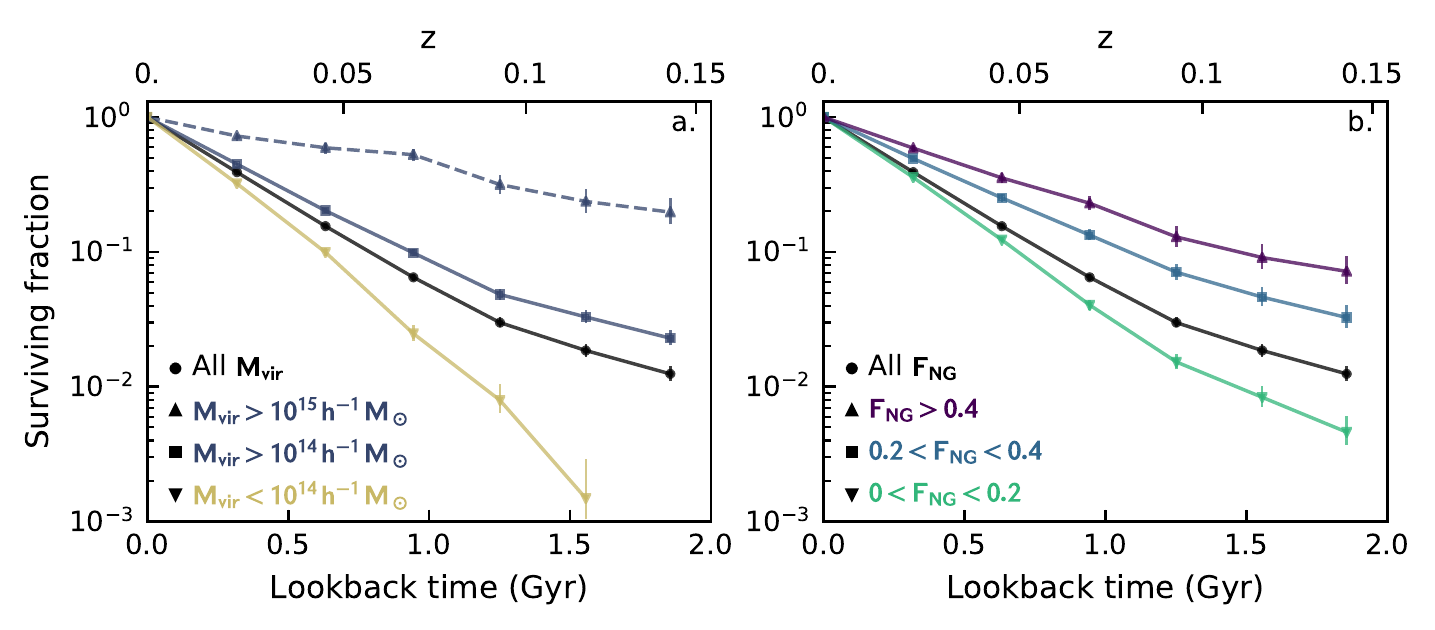}
    \caption{Surviving fraction for $z=0$ unrelaxed clusters as a function of lookback time.  The surviving fraction is the fraction of $z=0$ unrelaxed clusters whose MMPs are also classified as unrelaxed ($p_{AD} < 0.05$) in subsequent redshift snapshots.  Once the MMP of a cluster is classified as relaxed ($p_{AD} \ge 0.05$) in a snapshot, then the cluster is considered to have not survived.  Errorbars correspond to 68 per cent statistical errors estimated from the beta distribution following \citet{cameron2011}.  \textit{Left:} Divided by parent halo mass.  \textit{Right:}  Divided by $F_\mathrm{NG}$ (see Equation~\ref{eq:f_unrel}).}
    \label{fig:survive}
\end{figure*}

With the available redshift snapshots and merger trees we can probe how long $\mathrm{NG_{1D}}$ systems remain in the NG state.  Given that $\mathrm{NG_{1D}}$ clusters are associated with transient events like MMs (see section~\ref{sec:MMtime}), it is interesting to estimate the characteristic time required for clusters, on average, to return to a relaxed dynamical state.  To do this we trace the MMP of each redshift zero cluster back through the merger trees out to a given redshift.  We then keep all $z=0$ $\mathrm{NG_{1D}}$ clusters whose MMPs have at least ten member galaxies in all snapshots back to this redshift, which ensures that we can apply the AD test to the cluster MMPs in each snapshot.  We note that for this part of the analysis we consider one-dimensional velocities measured along the $z$-axis of the simulation box as opposed to the one random line-of-sight used up until this point.  The reason is that this is a simple way to ensure that we are measuring velocities along the same axis for clusters and all of their MMPs as we trace them back through the merger trees.
\par
To probe the rate at which NG $z=0$ systems cease to be classified as NG, we consider the survival curve\footnote{Survival curves are commonly used in radiobiology to determine the fraction of surviving cells as a function of radiation dose \citep[e.g.][]{deacon1984}.} for $\mathrm{NG_{1D}}$ clusters.  We measure the fraction of $\mathrm{NG_{1D}}$ systems at $z=0$ that ``survive" as we move to simulation snapshots at higher redshift.  We consider an $\mathrm{NG_{1D}}$ system at $z=0$ to have survived out to a redshift, $z$, if the MMPs of that cluster are classified as NG ($p_\mathrm{AD} < 0.05$) in \textit{all} redshift snapshots between $z=0$ and $z$.  $\mathrm{NG_{1D}}$ clusters at $z=0$ whose MMPs are classified as G in some snapshot between $z=0$ and $z$ are considered to not have survived at redshift $z$.  These ``surviving fractions" give us a quantitative measure of how quickly the population of $\mathrm{NG_{1D}}$ clusters evolves back to the $\mathrm{G_{1D}}$ state.  We trace the MMPs of $z=0$ $\mathrm{NG_{1D}}$ systems back to the redshift where the surviving fraction of all $\mathrm{NG_{1D}}$ systems reaches $\sim 1$ per cent.  This corresponds to $z \sim 0.15$ or a lookback time of $\sim 2$ Gyr.
\par
In Fig.~\ref{fig:survive} we plot the surviving fractions of $z=0$ $\mathrm{NG_{1D}}$ clusters as a function of lookback time.  At a lookback time of 0.0 Gyr the surviving fraction is unity, by construction, and then the surviving fraction decreases toward higher redshift.  In Fig.~\ref{fig:survive}a the black line corresponds to the surviving fraction for all unrelaxed clusters, and the coloured lines correspond to subsamples of halo mass.  The solid yellow line shows the surviving fraction for all low-mass clusters ($M_\mathrm{vir} < 10^{14}\,\mathrm{M_\odot}$), the solid blue line shows the surviving fraction for all high-mass clusters ($M_\mathrm{vir} > 10^{14}\,\mathrm{M_\odot}$), and the dashed blue line shows the surviving fraction for \textit{very} high-mass clusters ($M_\mathrm{vir} > 10^{15}\,\mathrm{M_\odot}$).  The decline in surviving fraction depends on halo mass, with the surviving fractions for low-mass clusters falling off the most quickly and the fractions for the most massive halos declining at the slowest rate.  At a lookback time of $\sim$ 2 Gyr, the surviving fractions for the vast majority of halos are $\lesssim$ a few per cent.  Only the clusters with $M_\mathrm{vir} > 10^{15}\,\mathrm{M_\odot}$ have surviving fractions which persist above 10 per cent for longer than 1 Gyr.  Therefore clusters identified as NG in one-dimension at $z=0$ do not appear NG for long (on average), though the precise timescales depend on halo mass.
\par
In Fig.~\ref{fig:survive}b we now show surviving fraction divided by $F_\mathrm{NG}$ instead of halo mass.  A reminder that $F_\mathrm{NG}$ is the fraction of random projections along which a given group/cluster is classified as NG, therefore it is a measure of how unrelaxed a system is along many lines-of-sight as opposed to just one.  In Fig.~\ref{fig:survive}b the green line corresponds to $0 < F_\mathrm{NG} < 0.2$, the blue line corresponds to $0.2 < F_\mathrm{NG} < 0.4$, the purple line corresponds to $F_\mathrm{NG} > 0.4$, and the black line corresponds to all values of $F_\mathrm{NG}$ (same line as in panel a.).  A clear trend is visible, where clusters with the lowest values of $F_\mathrm{NG}$ also have the lowest survival fractions.  As $F_\mathrm{NG}$ increases so does the survival fraction.  This is expected as subsamples with high $F_\mathrm{NG}$ are samples of NG clusters with high purity (ie. fewer clusters misidentified as NG due to projection).  We emphasize that while Fig.~\ref{fig:survive}a \& b are not independent since $M_\mathrm{vir}$ and $F_\mathrm{unrel}$ are correlated (see Fig.~\ref{fig:Mvir_Funrel}).  However, throughout this paper we continue to see trends with $F_\mathrm{unrel}$ at fixed cluster mass and vice versa.
\par
We can define a ``half-life" for $z=0$ $\mathrm{NG_{1D}}$ clusters to be the lookback time at which point the surviving fraction is equal to 50 per cent.  For the total population this half-life is roughly 0.5 Gyr, considering subsamples with high halo-mass or high purity (high $F_\mathrm{NG}$) extends this half life up to $\sim$1 Gyr.  The fact that low-mass clusters have very low surviving fractions is consistent with our finding that many low-mass $\mathrm{NG_{1D}}$ clusters are seemingly quite relaxed and just misidentified due to projection (Fig.~\ref{fig:Mvir_Funrel}).  We reiterate that whether or not an $\mathrm{NG_{1D}}$ cluster survives is based on measurements of the one-dimensional velocity distribution, which is analogous to what observers measure for galaxy clusters.

\section{Discussion \& conclusions}
\label{sec:discussion}

In this study we identify G and NG galaxy clusters in a large dark-matter only simulation using an observational technique based on the one-dimensional cluster velocity profile.  The main objective of this work is to test how well the one-dimensional Anderson-Darling test is able to identify physical differences between clusters halos.  By classifying $\mathrm{G_{1D}}$ and $\mathrm{NG_{1D}}$ clusters using observational methods, we can directly compare the results of this work to observed clusters.  The main results of this work are the following:

\begin{enumerate}[label={\arabic*.}, itemsep=0.5em, leftmargin=1em]
	\item Time since last major merger is systematically shorter for $\mathrm{NG_{1D}}$ systems compared to $\mathrm{G_{1D}}$ systems (Fig.~\ref{fig:majormerger}).  This difference is strongest for high-mass clusters, whereas little difference is seen for lower mass systems.
    
    \item The time since infall (onto the present-day parent halo) is systematically shorter for satellites in $\mathrm{NG_{1D}}$ systems relative to $\mathrm{G_{1D}}$ systems (Fig.~\ref{fig:infallTime}).
    
    \item The non-Gaussianity of high-mass cluster velocity profiles is due to both major mergers as well as minor mergers and the accretion of isolated galaxies.  However, for low-mass clusters the non-Gaussianity seems to trace minor mergers and isolated accretion and not major mergers (Fig.~\ref{fig:MMtime_pFrac} and \ref{fig:infallTime_pFrac}).  
    
    \item The fraction of $\mathrm{NG_{1D}}$ clusters increases as a function of both halo mass and redshift.  The stronger increase is with halo mass, while the proportion of $\mathrm{NG_{1D}}$ systems increases more modestly with redshift (Fig.~\ref{fig:NGfrac_z}).
    
    \item On average, $\mathrm{NG_{1D}}$ systems remain NG for 0.5-1 Gyr (Fig.~\ref{fig:survive}).
    
    \item The difference between G and NG systems becomes much stronger when using three dimensional information to construct a sample of NG clusters with higher purity.  This suggests that the intrinsic dependencies of galaxy and cluster properties on dynamical state are likely underestimated observationally due to only having access to projected positions and velocities (Fig.~\ref{fig:MMtime_pFrac} and \ref{fig:infallTime_pFrac}).
\end{enumerate}

\subsection{Implications for galaxy quenching}

It is possible to use these results to interpret observational trends with cluster dynamic state.  For example, previous works have established that galaxies in NG systems tend to show signatures of being a relatively blue, star-forming, and active population compared to G systems \citep{ribeiro2010, hou2012, roberts2017}.  It is possible that these differences are related to differences in time-since-infall.  The fact that galaxies in NG clusters have been exposed to a dense environment for less time would naturally give rise to a galaxy population which is preferentially blue and star-forming relative to galaxies in G systems, without the need to invoke any specific quenching mechanism.  In an upcoming paper, we plan to use the infall history extracted from these simulations along with a quenching model to test whether differences in time-since-infall are sufficient to explain the dependence of star-forming fraction on cluster dynamical state observed in \citet{roberts2017}.  A second possibility is that these observed differences are related to physical differences between the halos of relaxed and unrelaxed clusters.  For example, unrelaxed clusters may have underdense and disturbed ICMs, which can affect the efficiency with which the cluster is able to environmentally quench satellites.  For example, \citet{roberts2016} show that X-ray underluminous systems show signatures of disturbed dynamics (see also, \citealt{popesso2007a}) and also host an excess of star-forming galaxies.  Environmental quenching mechanisms which involve interactions between galaxies and the ICM, such as ram pressure stripping or starvation, may therefore be less efficient in such systems.

\subsection{Estimating dynamic state along a single line-of-sight}

Observations of galaxy cluster dynamics are unavoidably restricted to one-dimensional line-of-sight velocity measurements, which is why we focus the majority of this analysis on NG clusters identified only in one dimension.  However, working with simulation data allows us to analyze a more pure sample where clusters have NG velocity distributions in a large fraction of random cluster projections.  We find that the separation between properties of G and NG clusters are consistently enhanced when considering a sample of NG clusters with a higher purity compared to the one-dimensional case.  This is due to the fact that observationally we only have access to one line-of-sight, and the fact that a cluster looks unrelaxed along one, random, line-of-sight is not enough to say conclusively that a given cluster is unrelaxed on the whole.  Indeed, many of the simulated clusters in this work which appear NG along one random projection, show little evidence for disturbed dynamics along other projections.
\par
In some sense this is discouraging, as the three-dimensional information required to more accurately classify cluster dynamical state is not accessible observationally.  On the other hand, the fact that we still see systematic differences in cluster properties such as time since MM and satellite time since infall, between G and NG clusters identified in one-dimension is encouraging.  These differences demonstrate that given a large enough sample, NG clusters identified in one-dimension are indeed preferentially unrelaxed relative to G clusters, despite the sample impurity.  As a result of this observational impurity, the differences which have been observed between large samples of G and NG clusters \citep{hou2009, ribeiro2010, carollo2013, ribeiro2013a, roberts2017, costa2018, roberts2018, nascimento2019} are almost certainly lower-limits to the true, underlying dependencies of cluster properties on dynamical state.  On a system-by-system basis, this impurity suggests that simply classifying the one-dimensional velocity profile is not enough to classify the underlying dynamical state with confidence.  To get a more comprehensive picture of cluster dynamics on a system-by-system basis, it is more useful to combine other dynamic probes alongside line-of-sight velocities, such as: X-ray morphology \citep[e.g.][]{roberts2018}, BCG offsets \citep[e.g.][]{lopes2018}, magnitude gaps \citep[e.g.][]{lopes2018}, galaxy spatial distributions \citep[e.g.][]{wen2013}, velocity dispersion profiles \citep[e.g.][]{bilton2018}, and more.  Many of these observational relaxation proxies are easily derived for groups and clusters in large redshift surveys (excluding X-ray proxies), and therefore identifying samples of unrelaxed systems using many observational tests will help mitigate some of the inherent uncertainty of individual probes.

\section*{Acknowledgements}
We thank the anonymous referee for their detailed comments which have significantly improved the manuscript.  IDR and LCP are supported by the National Science and Engineering Research Council of Canada.  This work was made possible thanks to a large number of open-source software packages, including:  \texttt{AstroPy} \citep{astropy2013}, \texttt{Matplotlib} \citep{hunter2007}, \texttt{NumPy} \citep{vanderwalt2011}, \texttt{Pandas} \citep{mckinney2010}, \texttt{SciPy} \citep{jones2001}, \texttt{Topcat} \citep{taylor2005}, \texttt{ytree} \citep{smith2018}.
\par
The CosmoSim database used in this paper is a service by the Leibniz-Institute for Astrophysics Potsdam (AIP).  The MultiDark database was developed in cooperation with the Spanish MultiDark Consolider Project CSD2009-00064.  The authors gratefully acknowledge the Gauss Centre for Supercomputing e.V. (www.gauss-centre.eu) and the Partnership for Advanced Supercomputing in Europe (PRACE, www.prace-ri.eu) for funding the MultiDark simulation project by providing computing time on the GCS Supercomputer SuperMUC at Leibniz Supercomputing Centre (LRZ, www.lrz.de).

%%%%%%%%%%%%%%%%%%%%%%%%%%%%%%%%%%%%%%%%%%%%%%%%%%



\bibliographystyle{mnras}
\bibliography{mnras_template} % if your bibtex file is called example.bib


%%%%%%%%%%%%%%%%%%%%%%%%%%%%%%%%%%%%%%%%%%%%%%%%%%

%%%%%%%%%%%%%%%%% APPENDICES %%%%%%%%%%%%%%%%%%%%%

%\appendix

%\section{Assigning cluster members in projection}

%%%%%%%%%%%%%%%%%%%%%%%%%%%%%%%%%%%%%%%%%%%%%%%%%%

% Don't change these lines
\bsp	% typesetting comment
\label{lastpage}
\end{document}